\documentclass[14pt]{article}

\usepackage{listings}  
\usepackage[utf8]{inputenc}

\usepackage[english]{babel}

\usepackage{amsmath}

\usepackage{caption}
\usepackage{graphicx}
\graphicspath{{main/}}
\DeclareGraphicsExtensions{.pdf,.png,.jpg,.EPS}
\usepackage[unicode, pdftex]{hyperref}
\usepackage{geometry}
\usepackage{subfig}

\usepackage{times}

\usepackage{blindtext}
\usepackage{hyperref}
\usepackage{subcaption}
\usepackage{multirow}
\usepackage{microtype} 

\begin{document}
\rm
\let\sc=\sf
\begin{center}
\LARGE \bf {Veiling of Photosphere Lines in the Spectra of UX ORI Stars
at Deep Light Minima. I. The Star RR Tau}\\
\vspace{10mm} \large {P. O. Dimitrieva$^{a}$, L. V. Tambovtseva$^{b}$, V. P. Grinin$^{b*}$}
\end{center}
\vspace{0.6cm}
\normalsize
$^a$ \textit{St. Petersburg State University, St. Petersburg, 199034 Russia}\\
$^b$ \textit{Pulkovo Astronomical Observatory, Russian Academy of Sciences, St. Petersburg, 196140 Russia}\\
* \textit{e-mail: vgcrao@mail.ru}

\vspace{10mm}
\large {\bf Abstract}.\, The spectra of RR Tau star, which belongs to the family of young irregular variable UX Ori type stars, in its different brightness states have been studied using a comparative analysis. Selected spectra of the star that are obtained with the Nordic Optical Telescope at various times when its brightness ranged from $10.6^m$ to $13.9^m$ have been presented. The veiling of spectral lines at brightness minima by circumstellar emission has been considered, and its origin has been discussed.  \\


\section{INTRODUCTION}

The star RR Tau belongs to the UX Ori (UXOR) family of stars. Its parameters: the mass of $M_*$ = 2.5 $M_\odot$, the radius of  $R_*$ = 2.1 $R_\odot$, $T_{eff}$ = 9750 K, log g = 3.5 \cite{rost1999} and $v$ $sini$ = 140 km/s \cite{five} are typical for Herbig Ae stars, to which most UX Ori type stars belong. Observational signs that can be used to classify stars as UX Ori include the presence of deep Algol-like minima, a bluing effect, and an increase in the degree of linear polarization in the minima. The brightness minima are caused by the shielding of stars by opaque fragments (clouds) of protoplanetary disks
tilted at a small angle to the line of sight \cite{graph}.

There are two possible different sources of spectral variability. One of them is typical for many young stars. In this case, the variability is due to variations in temperature and density in the emitting region and
along the line of sight. Observations show that such variations can be sufficiently strong and lead to strong variability of spectral lines. Another mechanism of spectral variability is characteristic only of UX Ori
type stars. It is caused by the coronagraphic effect created by dust components of the protoplanetary disk, which obscure the star from the observer and cause deep minima. By shielding the star from the observer,
the opaque fragments of the disk also shield that part of the gas shell in front of the star that is responsible for the absorption component of the line profile. As a result, central absorption is reduced and may disap
pear completely \cite{grinin1995}. This effect is clearly manifested
by the change in the intensity of hydrogen emission
lines, most noticeable in the H$_\alpha$ line.

To interpret the observed spectral variability, Grinin et al. considered two main scenarios for the star’s obscuration \cite{rr}: (1) the dust screen rises vertically above the circumstellar disk and (2) the screen crosses the line of sight, moving in azimuth with the disk. In both cases, the emission region model consists of a compact magnetosphere and a magnetocentrifugal disk wind. Comparison with observations showed
that the first scenario explains well the variability of the radiation flux of the equivalent width, and the asymmetry of the H$_\alpha$ line during eclipses, while the second scenario explains them only partially. This suggests that, in the case of RR Tau, the main causes of eclipses are a structured disk wind and/or charged dust raised along the field lines by the poloidal component of the magnetic field of the circumstellar disk

One of the important features of the spectra of young stars is the veiling effect: the blurring of the true (photospheric) spectrum by emission generated in the circumstellar environment. The problem of veiling of
photospheric spectra of young stars has long been discussed in the literature, but mainly only in relation to cool young T Tauri stars (see, for example, \cite{Nel}). The main source of veiling of the spectra of these stars are hot accretion spots on their surfaces. However, in UX
Ori type stars, the nature of the veiling is different and is associated with the coronagraphic effect that occurs when dust clouds transit across the star’s disk. As a result, photospheric lines become blurred or indistinguishable.

In our work, we study the spectral variability of RR Tau with an emphasis on the manifestations of veiling and its influence on the observed spectrum. The first steps in this direction were made in the study by Rogers et al. \cite{rodgers}, in which spectra of RR Tau with low spectral resolution were obtained both in the bright state and in deep brightness minima. They showed that the photospheric absorption lines of ionized iron
at minimum brightness were flooded with a little blue
shifted emission \cite{rodgers}.

In this paper, we continue our study of the nature of spectral line veiling in the spectrum of this star based on observations obtained with higher spectral resolution using the Nordic Optical Telescope in different brightness states \cite{rr}. The lines of Ca II, Fe II, OII, Na I D2, and He I, which demonstrate strong emission veiling at the brightness minima of the star, are considered. The kinematic conditions in the
region of formation of these lines are determined by the veiling emission.

\section{OBSERVATIONAL DATA}
The observations were carried out using the NOT (Nordic Optical Telescope), La Palma, Spain. The diameter of the primary mirror of NOT is 2.56 m. Spectral data for the study were obtained from observations under the Target-of-Opportunity (ToO) program \cite{not2010}. A fiber-optic echelle spectrograph (FIES) was used for spectral observations. It has the highest spectral resolution of  $R \sim 67000$, but for the RR Tau
observations, the lower resolution of $R \sim 25000$ was used based on the signal-to-noise ratio. 

The photometric data used in this paper are available thanks to observers from the Swedish Association for Amateur Astronomy (SAAF). A more detailed
description of the surveillance strategies and coopera
tion with SAAF are described in \cite{rr}.

Forty-five spectra were acquired for RR Tau between February 2019 and January 2023. The nominal exposure time was 1800 s. The wavelength coverage
is approximately 3700–9100 \AA \,. A range of star brightness variation: 
$V = 10.6^{m} - 13.9^m$.  The results of the observations are published in \cite{rr}. 

\section{ VEILING OF PHOTOSPHERIC LINES}
In Figs. 1–4, spectral lines, in which a strong veiling effect is observed at brightness minima, are shown. The averaged four spectra in the bright state (at $V = 10.58, 10.57, 10.64$ and $10.95^m$) are shown in red. Individual profiles differ little from the average ones. Average brightness value in bright state is $V= 10.7^m$. The same sections of the spectra at minimum brightness are shown in blue at  $V= 13.8^m$. The
observed spectra are given in the star’s coordinate system at radial velocity of $ v_r= +11$ km/s \cite{five}. The line intensities are normalized relative to the continuum for each spectral order and for each stellar brightness value.

Hydrogen lines of the star RR Tau, which exhibit strong emission in deep minima, are presented by Grinin et al. in \cite{rr}. Some spectral lines of RR Tau are presented in \cite{tamb2025}.

In the bright state of the star, the Ca II 3933 \AA\quad line has a deep photospheric profile. The components of the Ca II 8498 \AA\quad and Ca II 8662 \AA\quad triplet lines exhibit an asymmetric photospheric profile reminiscent of the inverse P Cygni profile shifted to the red side, which indicates gas accretion. At the minimum of the star, in the lines of the Ca II triplet (Fig. 1), asymmetry is observed and a sufficiently strong emission appears, which veils the photospheric absorption lines. The wings of the emission lines extend to  $\pm$200 km/s. The emission in the Ca II line of 3933 \AA\quad is bifurcated with the distance between each component and the line center being $ \sim $ 40 km/s.

\begin{figure}[htbp]
    \centering
    \subfloat{\includegraphics[width=0.33\linewidth]{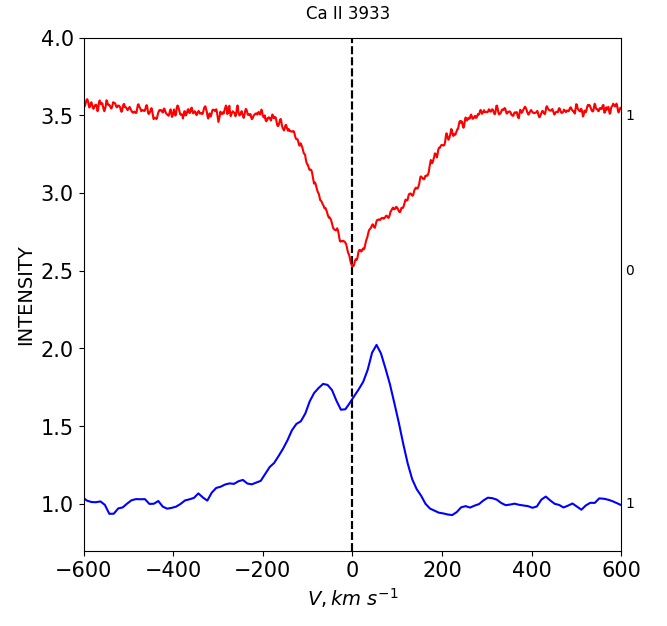}}\hspace{-0.1cm}
    \subfloat{\includegraphics[width=0.33\linewidth]{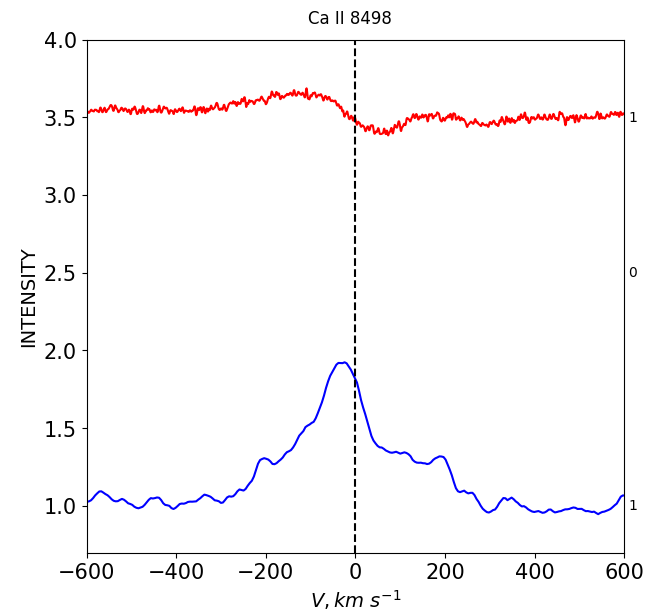}}\hspace{-0.1cm}
    \subfloat{\includegraphics[width=0.33\linewidth]{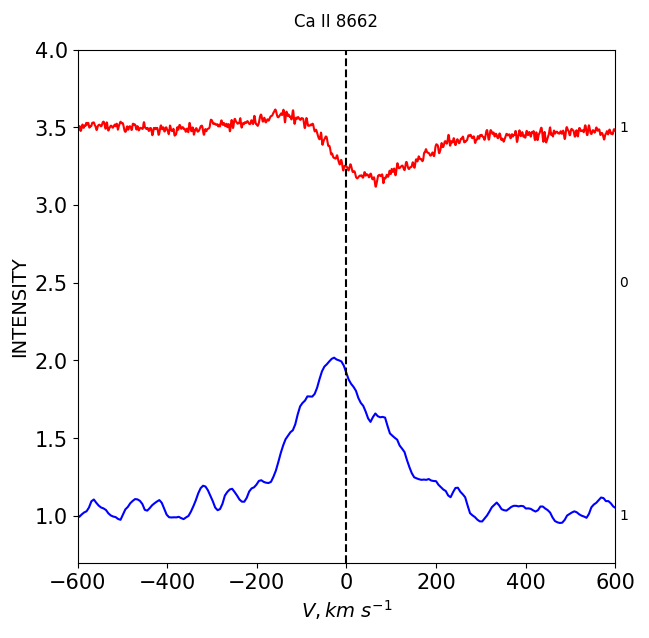}}
    \caption{Spectral lines of Ca II in the bright state of the star (red line) and in the deep minimum brightness (blue line). To reduce
noise in all Ca II spectra at minimum brightness, a weighted average operation was performed.}
    \label{fig:Ca1}
\end{figure}

\begin{figure}[htbp]
    \centering
    \subfloat{\includegraphics[width=0.33\linewidth]{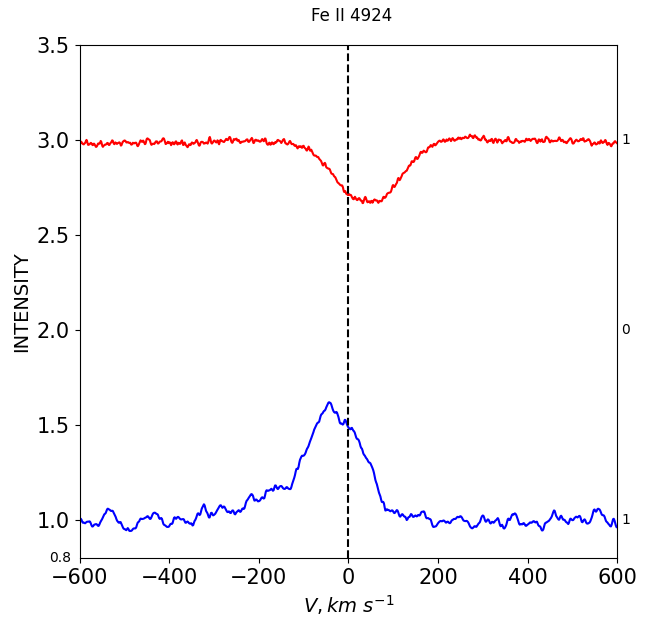}}\hspace{-0.1cm}
    \subfloat{\includegraphics[width=0.33\linewidth]{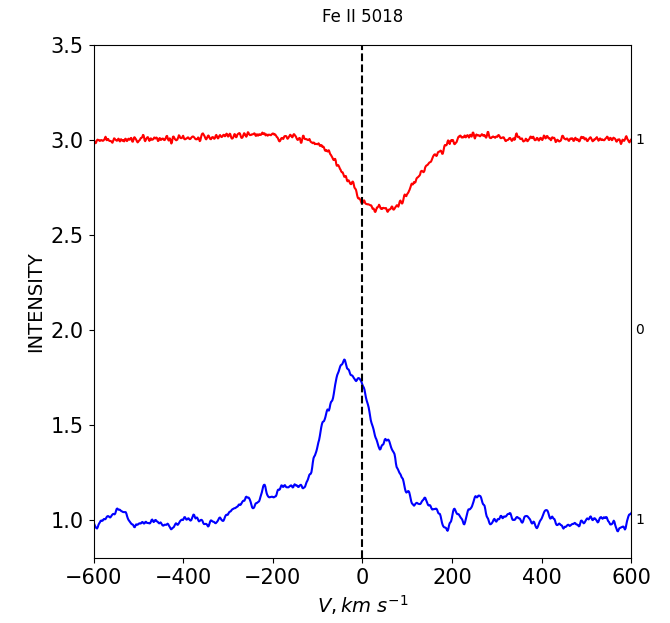}}\hspace{-0.1cm}
    \subfloat{\includegraphics[width=0.328\linewidth]{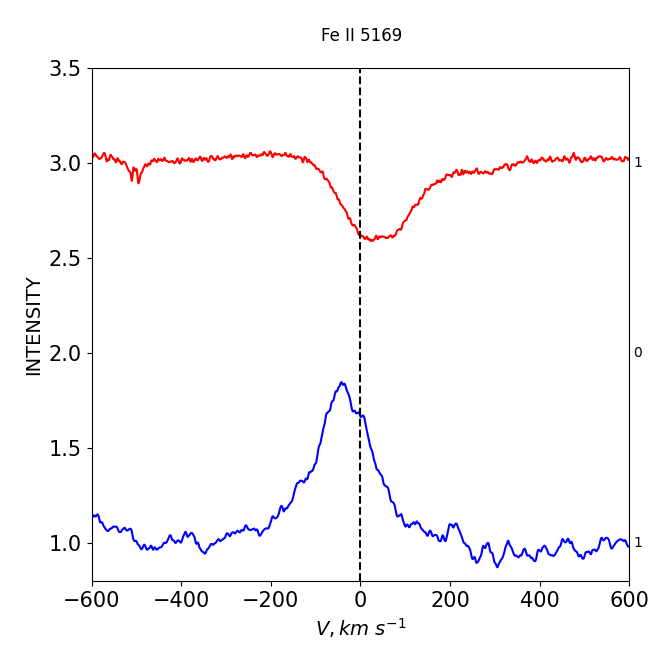}}
    \caption{Spectral lines of the Fe II triplet in the bright state (red line) and in the deep minimum (blue line).}
    \label{fig:Fe1}
\end{figure}

\begin{figure}[htbp!]%
    \centering
    \subfloat{{\includegraphics[width=0.41\linewidth]{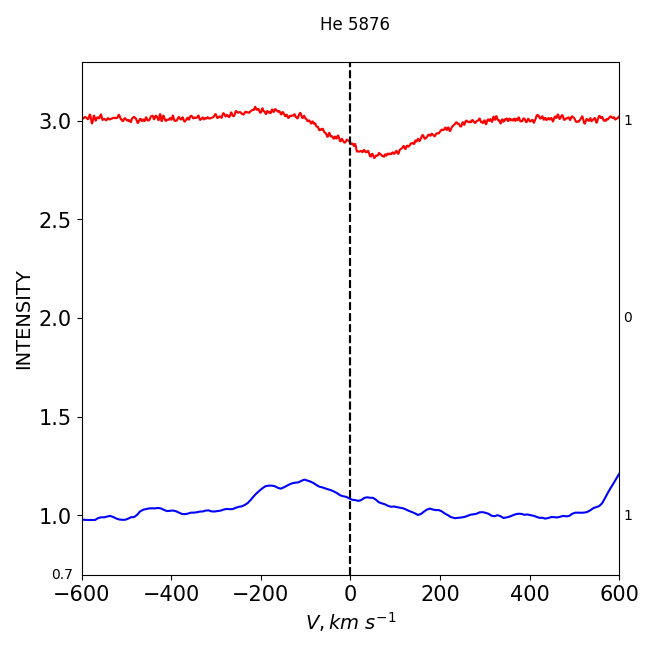} }}%
    \qquad
    \subfloat{{\includegraphics[width=0.41\linewidth]{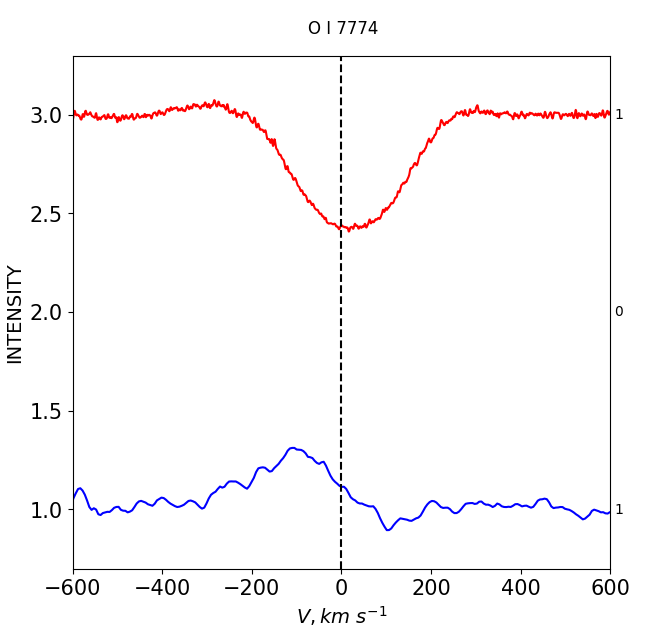} }}%
    \caption{Spectral lines of He I 5876  \AA \, and O I 7774  \AA \,. To reduce the noise in the spectrum at minimum brightness, a weighted
average operation was performed. }%
    \label{fig:He}%
\end{figure}

A similar pattern is observed for the Fe II triplet lines. In the bright state, photospheric lines are observed in all lines of the triplet that are little shifted to the red side (Fig. \ref{fig:Fe1}). In a weak state of the star, a
strong, little asymmetric emission appears, its maxima are shifted to the blue side by approximately 20 km/s. It completely obscures the absorption line.
In the bright state, the absorption lines of He I 5876 \AA \, and O I 7774 \AA \, are broadened by differential motions of the emitting gas (Fig. \,\ref{fig:He}). The absorption profile of the He I 5876 \AA \, line disappears at minimum brightness and weak emission appears.

The O I 7774 \, \AA \, line has a clearly defined profile when the star is in a bright state. At the minimum, the absorption profile of the line completely disappears and a weak asymmetric emission shifted to the blue
side appears.

In the bright state, the Na I doublet lines exhibit (see Fig. 4) narrow absorption components little shifted to the red side. At minimum brightness, emission, the peak of which is shifted to the blue side by
approximately 35 km/s appears, while narrow absorption components of the lines are preserved at the same velocities corresponding to the radial velocity of the star.

\begin{figure}[htbp!]%
    \centering
    \subfloat{{\includegraphics[width=0.5\linewidth]{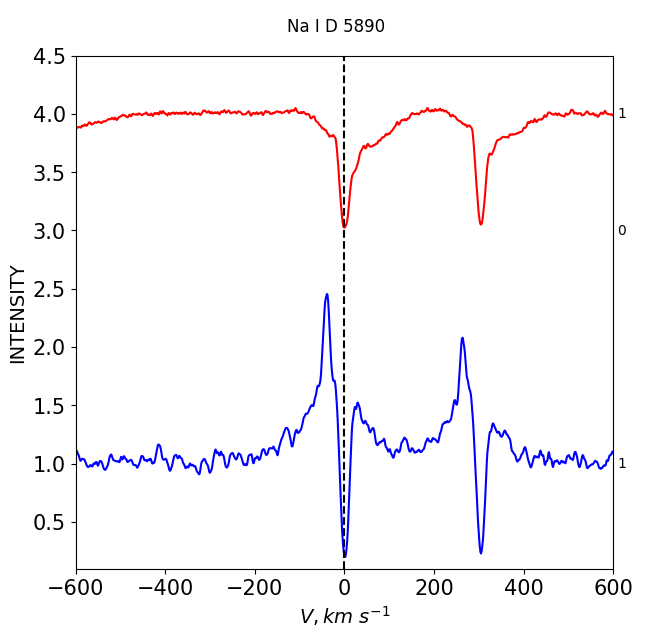} }}%
    \qquad
    \subfloat{{\includegraphics[width=0.42\linewidth]{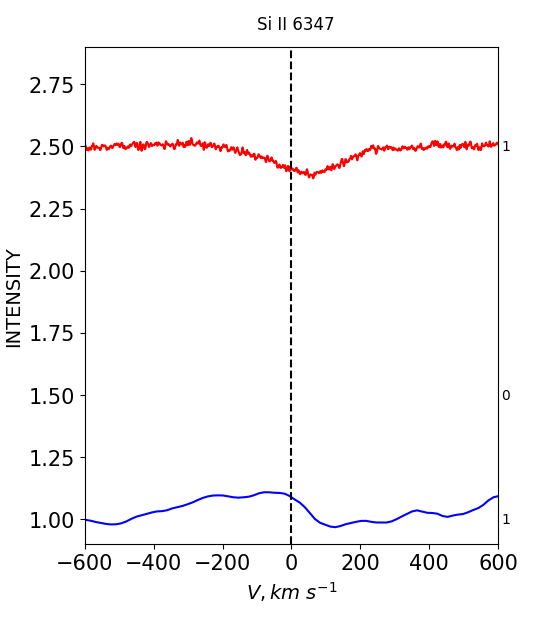} }}%
    \caption{ Spectral lines of Na I D and Si II 6347  \AA \,. To reduce noise in the Si II line at minimum brightness, a weighted average
operation was performed.}%
    \label{fig:Na1}%
\end{figure}

In the bright state, the Si II line has a photospheric absorption profile. At minimum brightness, a small blue-shifted emission, which obscures the photospheric line, can be seen.

\section{RESULTS}
In the paper by Grinin et al. \cite{rr}, it is shown that the disk wind plays a dominant role in the formation of Balmer lines in the spectrum of RR Tau. In the bright state of the star, a little asymmetrical two-peak profile is observed in the H$_\alpha$ line. As the star enters the minimum, the intensity ratio of the blue and red peaks of the profile remains virtually unchanged or changes very little. For H$_\beta$ and H$_\gamma$ lines, besides the disk wind, magnetospheric accretion plays a significant role. These lines are formed in a less extended region around the star.

The Ca II, Fe II, and Si II lines are formed in the atmosphere of the star and in the accretion circumstellar region \cite{tamb2025}. In the bright state, the Ca II (8498/8662 \AA) and Fe II (4924/5018/5169 \AA) triplet
lines exhibit an absorption profile shifted to the red side. The observed shift indicates magnetospheric accretion. The shift may also be caused by additional emission that appears during the minimum. At the minimum brightness of the star, these lines exhibit emission that is little shifted to the blue side, which indicates the outflow of matter in the disk wind.

The He I 5876 \AA \  line is formed in the falling gas flow of the star’s magnetosphere, in the accretion spot at an electron gas temperature of at least 17000 K \cite{1999}. The disappearance of the absorption line at the minimum is due to the fact that, during an eclipse, the dust screen partially or completely covers the region of formation of the helium absorption line.

A similar pattern is observed in the O I 7774 \AA\ line. This line is formed in hot gas, in the magnetosphere, in the region, where gas falls onto the star. Its profile is shifted to the red side during moments when the star is in a bright state. At the minima, the absorption profile
of the line completely disappears and a weak emission shifted to the blue side appears. During an eclipse, an opaque dust screen covers the region of the magnetosphere, where this line is formed. A small contribution
to emission can be made by scattered light \cite{rr}.

The sodium resonant doublet is one of the most interesting spectral features of UX Ori stars. The Na I line doublet is formed in the star’s magnetosphere. The observed red-shifted absorption arises in the
accreting gas of the magnetosphere. At the minimum, the emission components of the lines are little shifted to the blue side, which indicates the outflow of matter. The source of the observed radiation may be the peripheral regions of the disk wind, which are not obscured by the dust screen \cite{tamb2025}. This is evidenced by the weak asymmetry of radiation and narrow interstellar lines. The central narrow absorption is formed in the circumstellar and interstellar medium \cite{rr}. Sodium
resonance lines provide clear evidence that gas motions in the vicinity of UX Ori-type stars occur in a complex manner, with infall and outflow of matter occurring simultaneously.

\section{CONCLUSIONS}

\quad In this paper, the spectral variability of the star RR Tau in different brightness states has been analyzed based on echelle spectra obtained using the Nordic Optical Telescope \cite{rr}. It has been traced in which region the observed spectral lines are formed and what
is the reason for their veiling, i.e., the blurring of the true spectrum by emission.

It has been shown that at brightness minima the photospheric absorption lines become weaker, while the emission components become stronger, which are most noticeable in the Ca II and Fe II lines. Similar spectral changes have been previously observed in the T Tauri RW Aur star during a deep and prolonged minimum of brightness \cite{takami},\cite{fecchini}. Spectral variability during deep fadings has also been observed in the well-known T Tauri AA Tau star \cite{bouvier} and in V409 Tau, a star similar to AA Tau \cite{rodriguez}. The cause of the veiling is circumstellar emission in the corresponding lines.

The He I 5876 \AA \quad and O I 7774 \AA \quad lines are formed in the accreting hot gas flow of the magnetosphere; during eclipses, they almost completely disappear as a result of screening by the dust screen.

The Na doublet lines are formed mainly in the star’s magnetosphere and partly in the disk wind. The observed variability is caused by shielding by dust clouds of both the star itself and its magnetosphere.
The emission in the lines at minimum brightness is formed mainly in the region of the disk wind that is external to the dust screen and is almost invisible in the bright state of the star. A model of this region has been
calculated for another star of this family, UX Ori \cite{15}.
It extends from the dust sublimation zone (0.4 AU) to approximately 3 AU and is not obscured by the dust shield. A single H$_\alpha$-line emission profile shifted to the blue side is formed in this region. The same type of
profile is observed in the metal lines at the brightness minima of RR Tau.

It is also worth noting a very interesting observational fact: the emission profiles of the sodium lines in the spectrum of RR Tau at brightness minima are very similar to the lines of the sodium doublet in the spectrum of the photometrically inactive Herbig Ae AB
Aur star \cite{bohm}. The only difference is that, in the spectrum of RR Tau, the sodium lines are somewhat blue shifted, indicating a slow outflow of matter in the peripheral region of the disk wind, whereas, in the
spectrum of AB Aur, these emissions do not show any shift. The indicated similarity of the emission profiles of sodium lines in the spectra of two Herbig Ae stars, whose disks are differently oriented relative to the
direction to the observer, deserves separate consideration and modeling. \\
\\
The authors thank the reviewer for helpful comments.

\newpage
\clearpage \vspace{1cm}


\medskip





\makeatletter
\renewcommand\@biblabel[1]{#1.} 
\makeatother

\renewcommand{\refname}{REFERENCES}


\end{document}